\documentclass[showpacs]{revtex4}
\usepackage{graphicx,psfrag,amsmath,amssymb,amsfonts,latexsym,color,dcolumn,
graphpap}
\begin{document}
\title{Neumann Casimir effect: a singular boundary-interaction approach}
\author{C. D. Fosco$^a$}
\author{F. C. Lombardo$^b$ }
\author{F. D. Mazzitelli$^b$ }
\affiliation{$^a$Centro At\'omico Bariloche and Instituto Balseiro,
 Comisi\'on Nacional de Energ\'\i a At\'omica, \\
R8402AGP Bariloche, Argentina} \affiliation{$^b$Departamento de
F\'\i sica {\it Juan Jos\'e Giambiagi}, FCEyN UBA, Facultad de
Ciencias Exactas y Naturales, Ciudad Universitaria, Pabell\' on I,
1428 Buenos Aires, Argentina.}
\date{today}

\begin{abstract}
Dirichlet boundary conditions on a surface can be imposed on a
scalar field, by coupling it quadratically to a $\delta$-like
potential, the strength of which tends to infinity. Neumann
conditions, on the other hand, require the introduction of an even
more singular term, which renders the reflection and transmission
coefficients ill-defined because of UV divergences. We present a
possible procedure to tame those divergences, by introducing a
minimum length scale, related to the non-zero `width' of a {\em
nonlocal\/} term. We then use this setup to reach (either exact or
imperfect) Neumann conditions, by taking the appropriate limits.
After defining meaningful reflection coefficients, we calculate the
Casimir energies for flat parallel mirrors, presenting also the
extension of the procedure to the case of arbitrary surfaces.
Finally, we discuss briefly how to generalize the worldline approach
to the nonlocal case, what is potentially useful in order to compute
Casimir energies in theories containing nonlocal potentials; in
particular, those which we use to reproduce Neumann boundary
conditions.
\end{abstract}
 \maketitle
\section{Introduction}\label{sec:intro}
Material bodies can modify the vacuum structure of a quantum field
theory, giving  rise to interesting physical phenomena, like forces
between neutral objects (Casimir effect) and changes in the decay
rates of excited atoms \cite{reviews}. In some cases, the effect of
the bodies can be grossly described by assuming that the fields
satisfy exact Dirichlet or Neumann boundary conditions on their
surfaces. This idealization, as well as the perfect conductor
approximation in QED, must of course be modified to cope with more
realistic situations. Indeed, inside a real conductor the fields do
not vanish. A more realistic description corresponds to a linear
relation between the field (and its derivative) on one side of the
conducting interface and the same objects on the other.

Boundary conditions are just an effective, approximate, macroscopic
way of taking into account the effects of the interaction between
the vacuum quantum fields and microscopic matter degrees of freedom
inside the bodies. A more refined way of taking that interaction
into account is to use linear response theory, whereby a generally
nonlocal, quadratic effective action for the quantum field is
obtained. The details about the microscopic interaction become then
subsumed  into the nonlocal kernels of this effective action term
\cite{prd08}.  This procedure does not, in general, yield exact
boundary conditions: on the one hand, the kernel is different from
zero on a finite width region. On the other, in a realistic
situation, it is a smooth bilocal function. Within the context of
Casimir physics, there are several reasons to consider this and
other kinds of `imperfect' versions of the exact Dirichlet and
Neumann boundary conditions. Firstly, phenomenology tells us that
realistic models for the electromagnetic properties of neutral
bodies can hardly be described by `sharp' boundary conditions on the
quantum fields. Secondly, the  perfect conductor approximation
presents difficulties, even from a purely theoretical standpoint:
divergences of the vacuum energy density close to the surfaces
\cite{candelas} and, as a consequence, some ambiguities and even
conceptual problems in the calculation of self-energies \cite{jaffe}
and its concomitant gravitational effects \cite{fulling}. Finally,
there may be  a practical reason to use a sharp but imperfect
version of a boundary condition: Dirichlet Boundary Conditions (DBC)
can be implemented numerically as interaction terms involving
surface deltas, taking the strong coupling limit at the end. The
analogue procedure is not known, however, for the case of Neumann
Boundary Conditions (NBC). Note that an alternative approach to Neumann 
boundary conditions, in terms of a special kind of ``matching conditions'' 
has been introduced in \cite{vassilevich}.

In this letter we study the problem of imposing (NBC) on a massless
real scalar field,  by means of boundary interaction terms. There
is, at first sight, a straightforward solution to this problem,
namely, to adapt the approach followed for the Dirichlet case, to
impose a different boundary condition.  That would imply to use
interaction terms proportional to the normal derivative of the
$\delta$-function. However, because these are highly singular
objects, the situation becomes more subtle. Indeed, as we have shown
in a previous paper \cite{derivatives}, in order to compute the
Casimir energy for potentials containing derivatives of the
$\delta$-function, it is necessary to introduce an ultraviolet
regulator.  We will show that the ultraviolet cutoff is needed to
compute the transmission and reflection coefficients for a single
mirror.  We shall introduce that cutoff explicitly into the theory,
by considering nonlocal interaction terms, showing also that the
cutoff can be naturally interpreted as the inverse of the width of
the mirror.

The structure of this letter is as follows: In Section
\ref{sec:boundary} we compute the reflection coefficients for
theories described by local and nonlocal potentials, emphasizing the
difficulties that the straightforward approach to the NBC has. We
then calculate, in Section~\ref{sec:casimir}, the Casimir energy for
flat mirrors using Lifshitz formula \cite{lifshitz}, both in the
local and nonlocal cases. We finally present, in Section
\ref{sec:general} the construction of nonlocal potentials for
mirrors of arbitrary shape, and also a possible generalization of
the worldline approach to cope with nonlocal potentials. This would
allow one, in principle, to extend the worldline approach
\cite{Gies} to the calculation of Casimir energies in the Neumann
case, by taking the appropriate limit.

\section{Boundary conditions and interaction terms}\label{sec:boundary}
Throughout this letter, we shall consider a massless real scalar
field $\varphi(x)$ in $d+1$ dimensions as the `vacuum' (as opposed
to matter) field. The aim of this section is to derive the boundary
conditions for the vacuum field from the knowledge of the
interaction term that accounts for its coupling to a mirror, by
solving its equations of motion. In particular, we want to explore
the nonlocal and NBC cases.

To that end, we first define the action $S$, which  will be assumed
to be of the form:
\begin{equation}\label{eq:defs}
S(\varphi) \;=\; S_0(\varphi) + S_I(\varphi)
\end{equation}
where the free action $S_0$ is given by
\begin{equation}\label{eq:defs0}
S_0(\varphi) = \frac{1}{2} \, \int d^{d+1}x \; \partial_\mu \varphi
(x)
\partial^\mu \varphi (x)
\end{equation}
while the term that implements the interaction with the mirror has
the structure
\begin{equation}\label{eq:defsi}
    S_I(\varphi) = - \frac{1}{2} \, \int d^{d+1}x \int d^{d+1}x'
    \varphi(x) V(x,x')\varphi(x') \;,
\end{equation}
with a real and symmetric kernel $V(x,x')$. The explicit form of the
kernel depends on the details of the interaction between the quantum
scalar field and the degrees of freedom in the mirror. The kernel
$V(x,x')$ is assumed to be invariant under translations in time
($x^0$) and in the parallel ($x^i, \, i=1,\ldots,x^{d-1}$) spatial
coordinates. Thus, denoting by $x_\parallel$ all the coordinates
except $x_d$, we assume that:
\begin{equation}\label{eq:vk}
V(x,x') \;=\;  V(x_\parallel-x'_\parallel;x^d,x'^d) \;.
\end{equation}
Then the equations of motion adopt the nonlocal form:
\begin{equation}\label{eq:eqofm}
    \Box \varphi(x_\parallel,x^d) \;=\; - \int d^{d+1}x'
    V(x_\parallel-x'_\parallel;x^d,x'^d)\varphi(x'_\parallel,x'^d) \;.
\end{equation}
Then we take advantage of translation invariance in $x_\parallel$ to
use the Fourier transformed version of (\ref{eq:eqofm}):
\begin{equation}\label{eq:eqofm1}
    (\partial_d^2 + k^2) \, {\widetilde\varphi}(k,x^d)
    \;=\; \int dx'^d \, {\widetilde V}(k;x^d,x'^d) \, {\widetilde\varphi}(k,x'^d) \;,
\end{equation}
where we introduced $k^\alpha$,  $\alpha = 0,1,\ldots,d-1$, and:
\begin{eqnarray}\label{eq:defou}
    \varphi(x_\parallel,x^d) &=& \int \frac{d^{d}k}{(2\pi)^d}
    \,e^{i k \cdot x_\parallel} \,{\widetilde\varphi}(k,x^d)
 \nonumber\\
     V(x_\parallel - x'_\parallel,x^d) &=& \int \frac{d^{d}k}{(2\pi)^d}
    \,e^{i k \cdot (x_\parallel - x'_\parallel)} \,
    {\widetilde V}(k;x^d,x'^d) \;.
    \end{eqnarray}

We are interested in solutions that look like plane waves far from
the mirror, and we want to be able to extract from the solution the
reflection and transmission coefficients. It is then natural to
treat the system as a scattering problem, writing the solution by
means of the corresponding Lippmann-Schwinger (L-S) equation:
\begin{eqnarray}\label{eq:ls}
    {\widetilde\varphi}(k,x^d) &=&
    {\widetilde\varphi}^{(0)}(k,x^d) \nonumber\\
    &+&\int dx'^d \int dx''^d \, \Delta(k;x^d,x'^d)
    {\widetilde V}(k;x'^d,x''^d) {\widetilde
    \varphi}(k,x''^d) \;,
\end{eqnarray}
where ${\widetilde\varphi}^{(0)}$ is the (incident) free-particle
wave, solution of
\begin{equation}
    (\partial_d^2 + k^2) {\widetilde\varphi}^{(0)}
    (k,x^d) = 0 \;.
\end{equation}
and $\Delta$ is the retarded Green's function. We shall assume the
free-particle solution to correspond to a wave incident from $x^d <
0$, namely: ${\widetilde\varphi}^{(0)}(k,x^d) = e^{i k^d x^d}$,
where $k^d \equiv \sqrt{ (k^0)^2 - \sum_{i=1}^{d-1} (k^i)^2} = k \,
> \, 0$ and $k^0 > 0$, which are just the mass shell conditions.

On the other hand, the retarded Green's function satisfies:
\begin{equation}
    (\partial_d^2 + k^2) \Delta(k;x^d - x'^d)  = \delta(x^d -x'^d)\;,
\end{equation}
(with retarded boundary conditions) and may be written more
explicitly as follows:
\begin{eqnarray}
    \Delta(k; x^d - x'^d) &=& \int \frac{d\nu}{2\pi} \,e^{i \nu (x^d - x'^d)}
    \frac{1}{ - \nu^2 + (k^0 + i \eta)^2 - \sum_{i=1}^{d-1} (k^i)^2}
    \nonumber\\
    &=&
    - \frac{i}{2 k} \, e^{ i k \, |x^d - x'^d| }\;.
\end{eqnarray}

Let us now solve the L-S equation in some particular cases.

\subsection{Dirichlet-like boundary conditions}
As a warming-up exercise, we consider a local kernel $V_D$ which may
be used to impose Dirichlet like boundary conditions:
\begin{equation}\label{eq:dirichlet}
    {\widetilde V}_D(k,x^d,x'^d) \;\equiv \;
    \mu_0(k) \, \delta(x^d) \, \delta(x'^d) \;.
\end{equation}
Inserting this into (\ref{eq:ls}), we obtain:
\begin{equation}
    {\widetilde \varphi}(k,x^d) \;=\;{\widetilde \varphi}^{(0)} (k, x^d)
    \,+\, \mu_0(k) \, \Delta(k;x^d,0) \,
    {\widetilde \varphi}(k,0)  \;,
\end{equation}
whence we obtain, by evaluating the equation above at $x^d=0$:
\begin{equation}
    {\widetilde \varphi}(k,0) \;=\; \frac{1}{ 1 + \frac{i
    \mu_0(k)}{2 k}} \;,
\end{equation}
and, finally:
\begin{equation}
    {\widetilde \varphi}(k,x^d) \;=\; e^{ i k x^d} \, - \,
    \frac{\frac{i\mu_0(k)}{2 k}}{ 1 + \frac{i \mu_0(k)}{2 k}}
    e^{ i k |x^d|}\equiv
{\widetilde \varphi}(k,x^d) \;=\; e^{ i k x^d} \, + \,
    r(k)
    e^{ i k |x^d|}
 \,.
\label{rdirloc}
\end{equation}
When $\frac{|\mu_0(k)|}{2 k} \to \infty$, we see that ${\widetilde
\varphi}(k,x^d)=0$ for $x^d >0$, $r(k)=-1$,  the wave is perfectly
reflected and the field satisfies DBC at $x_d=0$. An interesting
particular case is $\mu_0(k)=\gamma k$, that produces a constant
reflection coefficient and DBC in the limit $\gamma\to\infty$. This
kind of potentials are generated by massless fermion fields confined
to the mirror \cite{plb08}.

In the general case, the reflection ($R$) and transmission ($T$)
coefficients are:
\begin{eqnarray}
R &=& \vert r(k)\vert^2 = \frac{\frac{|\mu_0(k)|^2}{4 (k^d)^2}}{ 1 +
\frac{|\mu_0(k)|^2}{4 k^2} - \frac{ {\rm Im}(\mu_0)}{k}} \nonumber\\
T &=& \vert 1+ r(k)\vert^2 =  \frac{1 -\frac{ {\rm Im}(\mu_0)}{k}}{
1 + \frac{|\mu_0(k)|^2}{4 k^2} - \frac{ {\rm Im}(\mu_0)}{k}} \;.
\end{eqnarray}
It is easy to check that $R+T=1$.

\subsection{Neumann-like boundary conditions}
In this case, we consider a kernel:
\begin{equation}\label{eq:neumann}
    {\widetilde V}_N(k,x^d,x'^d) \;\equiv \;
    \mu_2(k) \, \delta'(x^d) \, \delta'(x'^d) \;.
\end{equation}
Now we obtain the relation:
\begin{equation}
    {\widetilde \varphi}(k,x^d) \;=\;{\widetilde \varphi}^{(0)} (k,x^d)
    \,+\, \mu_2(k) \,\big[\frac{\partial}{\partial x'^d} \Delta(k;x^d,x'^d)\big]_{x'^d=0} \, {\widetilde
    \varphi}'(k,0)  \;,
\end{equation}
which yields ${\widetilde \varphi}(k,0) \;=\;{\widetilde
\varphi}^{(0)}(k,0)$ ($=1$), i.e., no effect on the value of the
incident wave at the mirror. On the other hand, by taking a
derivative with respect to $x^d$ above, and setting $x^d = 0$:
\begin{equation}
    {\widetilde \varphi}'(k,0) \;=\;
    \frac{{\widetilde \varphi}^{'(0)}(k,0)}{ 1 - \mu_2(k) D(k) } \;,
\end{equation}
where:
\begin{equation}
D(k) \equiv \big[ \frac{\partial^2}{\partial x^d \partial x'^d}
\Delta(k;x^d,x'^d)\big]_{x_d=0, x'^d=0} \;.
\end{equation}
This quantity is ill-defined. Indeed, we see that it is linearly
divergent in the UV (large momenta in the $x^d$ direction).
Introducing a momentum cutoff $\Lambda$, we see that its regularized
version, $D_{reg}(k,\Lambda)$, behaves as follows:
\begin{equation}
    D_{reg}(k,\Lambda) \sim - \frac{\Lambda}{\pi} \,-\,
i\frac{k}{2} \;.
\end{equation}
There is a very clear physical meaning in this cutoff: indeed, as we
shall show in the next subsection, it may be interpreted as due to a
finite width $\epsilon$ for the mirror. In particular, it may be
implemented by using a kernel similar to the one in this subsection,
but with the derivatives of the deltas replaced by one of its
approximants.

Keeping the cutoff $\frac{\Lambda}{\pi} \equiv \epsilon^{-1}$
finite, we find a relation involving the derivatives of the free and
exact field configurations:
\begin{equation}
    {\widetilde \varphi}'(k,0) \;=\;
    \frac{ {\widetilde \varphi}^{'(0)}(k,0)}{ 1 + \mu_2(k)  (\epsilon^{-1} + i\frac{k}{2})} \;.
\end{equation}
It is possible (and convenient)  in this context to hide the cutoff,
by relating it to a quantity with a more direct physical meaning,
playing the role of a renormalization condition. For example,
introducing the ratio:
\begin{equation}
    \alpha \equiv \big[\frac{{\widetilde \varphi}'(k,0)}{ {\widetilde
    \varphi}^{'(0)} (k,0)} \big]_{k\to 0} \;,
\end{equation}
we may write $\epsilon$ in terms of $\alpha$ and $\mu_2(0)$:
$\epsilon^{-1} = \mu_2^{-1}(0) (\alpha^{-1} -1 )$.

Then we may write the general solution for the field in terms of the
function $\mu_2$ and the constant $\alpha$:
\begin{equation}
    {\widetilde \varphi}(k,x^d) \;=\; e^{i k x^d} -
    r(k) \, {\rm sign}(x^d) e^{ i k |x^d|} \;,
\end{equation}
where:
\begin{equation}
r(k) \;\equiv\; \frac{\frac{i k \mu_2(k)}{2}}{ 1 \,+\,
\frac{\mu_2(k)}{\mu_2(0)} (\alpha^{-1}-1)   + \frac{i k
\mu_2(k)}{2}} = \frac{\frac{i k \mu_2(k)}{2}}{ 1 + \mu_2(k)
(\epsilon^{-1} + i\frac{k}{2})} \;. \label{rneuloc}
\end{equation}
It is clear that NBC emerge if $r(k) \to 1$, and this is the case
for an infinitesimal $\mu_2(k)=-\epsilon$. Indeed, writing
$\mu_2^{-1}=-\epsilon^{-1}+\Gamma^{-1}$ we have
\begin{equation}
r(k)=\frac {1}{1-\frac{2 i}{\Gamma k}}\, .
\end{equation}
Note that, in the limit $\Gamma k\gg 1$,  this corresponds to a
``soft'' NBC with  ${\widetilde \varphi}'(k,0)\sim \Gamma^{-1} $.

\subsection{Nonlocal kernel}
We consider here a case which includes the mirror's size into the
game, albeit not in the most general form. It does allow one,
however, to reach both the Dirichlet and Neumann cases as particular
limits. Besides, it automatically introduces a regularization
(related to a finite width) for the Neumann case.

This example corresponds simply to using the kernel:
\begin{equation}
    {\widetilde V}_\epsilon(k;x^d,x'^d) \;=\; \sigma(k) \,
    g_\epsilon(x^d) \, g_\epsilon(x'^d)
\label{nonlocalkernel}
\end{equation}
where $g_\epsilon(x^d)$ is a function localized on a region of size
$\epsilon$. We shall assume, for the sake of concreteness, its
support to be the interval $[-\epsilon/2,+\epsilon/2]$. It may even
depend on $k$: everything we shall do in what follows would remain
valid had one included such a dependence. The same holds true for an
eventual dependence of $\sigma$ on $\epsilon$. As already stressed,
one can think of the nonlocal kernel as coming from the integration
of microscopic degrees of freedom living on the mirror and
interacting with the quantum field $\varphi$. Although from this
point of view the form of the kernel given in (\ref{nonlocalkernel})
may be non realistic, it will be sufficient in order to show the
regularizing effect of a nonlocality in the normal direction.

We shall not use specific forms for the function $g_\epsilon(x^d)$
yet. However, one may think of size-$\epsilon$ approximants of the
$\delta$ function or of its derivative, although the results we
shall obtain will not depend on those assumptions.

The application of (\ref{eq:ls}) to this case yields:
\begin{eqnarray}\label{eq:lsnl}
    {\widetilde\varphi}(k,x^d) &=&
    {\widetilde\varphi}^{(0)}(k,x^d) \nonumber\\
    &+& \sigma(k) \, \left[\int dx'^d \, \Delta(k;x^d,x'^d)
    g_\epsilon(x'^d)\right] {\widetilde \varphi}_g(k) \;,
\end{eqnarray}
where ${\widetilde \varphi}_g(k) \equiv \int dx^d \, g_\epsilon(x^d)
\, {\widetilde \varphi}(k,x^d)$.

Multiplying both members of (\ref{eq:lsnl}) by $g_\epsilon(x^d)$ and
integrating over $x^d$, we find the relation:
\begin{equation}\label{eq:relnl}
    {\widetilde \varphi}_g(k) \;=\; \frac{{\widetilde
    \varphi}^{(0)}_g(k)}{ 1 - \sigma(k) \, \Delta_g (k,\epsilon)}
\end{equation}
where:
\begin{equation}\label{eq:deltagn}
\Delta_g (k,\epsilon) \;\equiv\; \int dx^d \int dx'^d
g_\epsilon(x^d)\Delta(k;x^d,x'^d)
 g_\epsilon(x'^d) \;.
\end{equation}
Note that this object is finite for approximants that are
square-integrable, although the limit when $\epsilon \to 0$ may be
singular. For example, to reproduce the Neumann case, one may
consider  the function:
\begin{equation}\label{eq:gneu}
    g_\epsilon(x^d) \;=\; - \frac{4}{\epsilon^2} \,
    \theta\left(\frac{\epsilon}{2} - \vert x^d\vert \right) \, {\rm sign}(x^d) \;,
\end{equation}
which yields a finite result for $\Delta_g$ whenever $\epsilon \neq
0$:
\begin{equation}\label{eq:deltagen}
\Delta_g(k,\epsilon) \,=\, \frac{32}{k^2 \epsilon^4} \, \left[
\frac{\epsilon}{2} + \frac{1}{2 i k} (e^{i k \epsilon} -1) -
\frac{2}{i k} (e^{i k \frac{\epsilon}{2}} -1) \right] \;.
\end{equation}

To proceed, we insert (\ref{eq:relnl}) into (\ref{eq:lsnl}),
obtaining:
\begin{eqnarray}\label{eq:lsnl1}
    {\widetilde\varphi}(k,x^d) &=&
    {\widetilde\varphi}^{(0)}(k,x^d) \,+\,
        \frac{ \sigma(k) \,{\widetilde \varphi}^{(0)}_g(k)}{ 1 -
\sigma(k) \, \Delta_g (k,\epsilon)}
    \nonumber\\
    & \times & \frac{1}{2 i k} \,  \int dx'^d \, e^{i k |x^d-x'^d|}
    \,g_\epsilon(x'^d) \;.
\end{eqnarray}
Then we evaluate the equation above for two different situations,
both corresponding to points outside of the mirror: either $x^d >
\frac{\epsilon}{2}$, or $x^d < -\frac{\epsilon}{2}$, what yields:
\begin{eqnarray}
    {\widetilde\varphi}_> (k,x^d)  &=& e^{i k x^d}
    \, \left[ 1 + \frac{ \frac{\sigma(k)}{2 i k} {\widetilde
    \varphi}^{(0)}_g(k) {\widetilde \varphi}^{*(0)}_g(k)}{1 -
    \sigma(k) \, \Delta_g(k,\epsilon) }\right]\equiv t(k) e^{i k x^d}
    \;,
    \nonumber\\
    {\widetilde\varphi}_< (k,x^d)  &=& e^{i k x^d}
    + e^{- i k x^d}\,  \frac{ \frac{\sigma(k)}{2 i k} {\widetilde
    \varphi}^{(0)}_g(k)
    {\widetilde \varphi}^{(0)}_g(k)}{1 - \sigma(k) \,
\Delta_g(k,\epsilon) } \equiv e^{i k x^d} + r(k) e^{-i k x^d} \;,
\label{rytneumann}
\end{eqnarray}
respectively. Note that ${\widetilde \varphi}^{(0)}_g(k)  = \int
dx^d e^{i k x^d} g_\epsilon(x^d)$, may be thought of as the Fourier
transform of $g_\epsilon$. Using the notation: $\int dx^d e^{- i k
x^d} g_\epsilon(x^d) \equiv {\tilde g}_\epsilon (k)$ to make that
property explicit we write the $T$ ant $R$ coefficients as follows:
\begin{eqnarray}
    T(k,\epsilon)  &=& \vert t(k)\vert ^2 = \, \left| 1 + \frac{ \frac{\sigma(k)}{2 i k}
    |{\widetilde g}_\epsilon (k)|^2}{1 - \sigma(k) \, \Delta_g
(k,\epsilon)}\right|^2
    \;,
    \nonumber\\
    R(k,\epsilon)  &=&  \vert r(k) \vert^2 = \left| \frac{ \frac{\sigma(k)}{2 i k}
[{\widetilde g}^*_\epsilon(k)]^2}{1 - \sigma(k) \,
\Delta_g(k,\epsilon)} \right|^2 \;.
\end{eqnarray}
After some calculations it is possible to show that $R+T=1$. At this
point it is worth to note that, generally speaking, nonlocality can
induce violations to this relation, since the interaction with
microscopic degrees of freedom in the mirror may affect unitarity.
However, this is not the case for nonlocal kernels of the form
(\ref{nonlocalkernel}).

For the Neumann-like case, we may consider the $\epsilon \to 0$
limit with the $g_\epsilon$ introduced in (\ref{eq:gneu}). This
yields for $\Delta_g$ the asymptotic behaviour:
\begin{equation}
\Delta_g(k,\epsilon) \, \sim \, - \frac{4}{3} \epsilon^{-1} - i
\frac{k}{2} \;\;\;,\; \epsilon \sim 0 \;,
\end{equation}
as it may be seen from (\ref{eq:deltagen}).

Note that, for the same approximant, we find:
\begin{equation}
{\widetilde g}_\epsilon(k) \;=\; - \frac{16}{i k \epsilon^2} \,
\sin^2(\frac{k \epsilon}{4}) \;. \label{gexplicit}
\end{equation}

To reach the exact NBC from the nonlocal case, one may use an
appropriate dependence of $\sigma$, namely, choosing a
$\sigma(k,\epsilon)$ such that that condition is approached in the
limit when $\epsilon \to 0$. To that end, we take the derivative of
(\ref{eq:lsnl1}) at the origin, and see that, when $\epsilon \sim
0$:
\begin{equation}\label{eq:lsnl2}
    {\widetilde\varphi}'(k,0) \simeq
i k \left[ 1 \,+\, \frac{k}{2 i} \frac{ \sigma(k,\epsilon)}{ 1 -
\sigma(k,\epsilon) \, \Delta_g (k,\epsilon)} \right]
\end{equation}
thus, using a $\sigma$ such that:
\begin{equation}
\frac{ \sigma(k,\epsilon)}{ 1 - \sigma(k,\epsilon) \, \Delta_g
(k,\epsilon)} = -\frac{2 i}{k} \;,
\end{equation}
yields NBC, i.e. $r(k)=1$ and $t(k)=0$. It is worth noting that, as
in the local case, the NBC are obtained for an infinitesimal value
of $\sigma$,  that is $\sigma\simeq -3\epsilon/4$.

Finally, we stress that the Dirichlet-like case is obtained by
considering $g_\epsilon$ to tend to a  $\delta$ function. This
implies ${\widetilde g}_\epsilon(k) \to 1$, and $\Delta_g(k\epsilon)
\to \frac{1}{2 i k}$, what (setting $\sigma \to \mu_0$) reproduces
the Dirichlet like case result.

\section{Casimir energy}\label{sec:casimir}
We will now compute the Casimir energy for the configuration of two
identical mirrors centered at $x_d=0$ and $x_d=a$ in $3+1$
dimensions.  In principle, the Casimir energy can be obtained from
the nonlocal effective action using path integral techniques
\cite{derivatives,kardar}. However, it is simpler to use Lifshitz
formula \cite{lifshitz}. Indeed, we just need the analytic
continuation to the imaginary frequency axis of the reflection
coefficients already computed in the previous section. The
reflection coefficients $r(k)$ depend on the wave vector
$k^\alpha=(k^0,k^1,k^2)$.  Denoting by $\bar r$ the analytic
continuation $\bar r=r(k_0=i\xi,k^1,k^2)$, according to Lifshitz
formula, the Casimir energy then reads
\begin{equation}
E(a)=\frac{1}{2\pi}\int_0^\infty
d\xi\int\frac{d^2k}{(2\pi)^2}\log(1-\bar r^2e^{-2\kappa a})\,\, ,
\label{lifshitz}
\end{equation}
where $\kappa=\sqrt{k_1^2+k_2^2+\xi^2}$.

It is easy to see that, due to the symmetries we are assuming, $\bar
r$ depends only on $\kappa$. Therefore, using spherical coordinates
in momentum space Lifshitz formula can be written as
\begin{equation}
E(a)=\frac{1}{4\pi^2}\int_0^\infty d\kappa\, \kappa^2 \log(1-\bar
r^2e^{-2\kappa a})\,\, , \label{lifshitz2}
\end{equation}
which is well defined as long as $\bar r^2<1$.

Let us now discuss the main properties of the reflection
coefficients, for the local and nonlocal interaction terms
previously considered. For the Dirichlet-like boundary conditions
considered in Section 2.1, the analytic continuation of the
reflection coefficient, which can be read from (\ref{rdirloc}), is
given by:
\begin{equation}
 \bar r^2 =\left(\frac{\mu_0(\kappa)}{2\kappa+\mu_0(\kappa)}\right)^2\, .
\label{rdir}
\end{equation}
Note that $\bar r$ is well defined and smaller than 1 for
$\mu_0(\kappa)>0$. Besides, it does reproduce the Dirichlet Casimir
energy in the limit $\mu_0(\kappa)/\kappa\rightarrow\infty$.
Inserting Eq.(\ref{rdir}) into Eq.(\ref{lifshitz2}) we reproduce the
usual result for $\delta$-potentials \cite{milton}.

For the local Neumann-like case described in Section 2.2, the
reflection coefficient is given in (\ref{rneuloc}), and its analytic
continuation reads
\begin{equation}
 \bar r^2 =
\left[\frac{\mu_2(\kappa)\kappa}{2+\mu_2(\kappa)(2\epsilon^{-1}+\kappa)}
\right]^2\; ,
\end{equation}
and it is well-defined except when $-\epsilon < \mu_2(\kappa)<0$.
The Neumann Casimir energy can be reproduced when $\mu_2\to
-\epsilon$ from below.

For the Neumann-like nonlocal boundary term considered in Section
2.3, equations (\ref{rytneumann}) and (\ref{gexplicit}) allow we to
derive
\begin{equation}
 \bar r^2 =\left[\frac{128\, \bar\sigma}{\epsilon^4 \kappa^3(1-\bar\sigma\bar\Delta_g)}
 \sinh^4(\frac{\kappa\epsilon}{4})\right]^2\, ,
 \label{rnonloc}
\end{equation}
where $\bar\sigma$ and $\bar \Delta_g$ denote the analytic
continuations of $\sigma$ and $\Delta_g$ respectively. In order to
reproduce Neumann boundary conditions, $\bar\sigma$ must be a rather
involved function of $\kappa$ and $\epsilon$. However, when
computing the Casimir energy for mirrors separated by a distance
$a\gg\epsilon$, we expect the relevant values of $\kappa$ to satisfy
$\kappa\epsilon\ll 1$. In this limit, that function simplifies to
$\sigma=-\frac{3}{4}\epsilon$.  If we assume that $\sigma$ is a
constant, the Casimir energy for this reflection coefficient is well
defined as long as $\sigma <-\frac{3}{4}\epsilon $ or $\sigma >0 $.
These properties are illustrated in Fig. 1.

\begin{figure}[h!t]
\centering
\includegraphics[width=9.5cm]{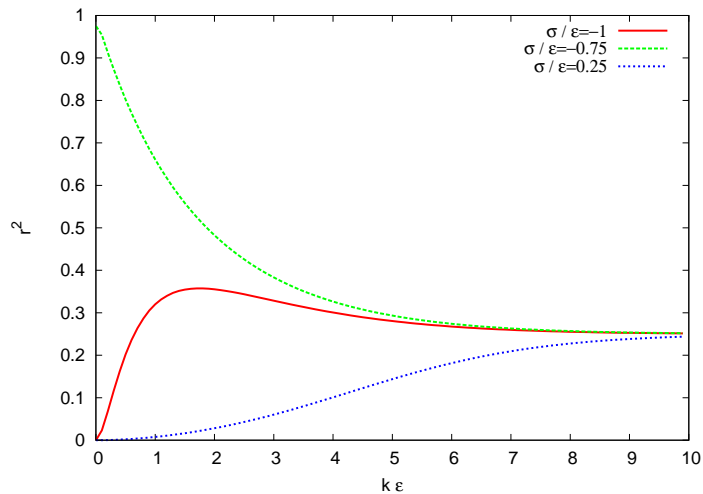}
\caption{The reflection coefficient given in (\ref{rnonloc}) as a
function of $x=\kappa\epsilon$, for different values of
$\sigma/\epsilon$. In the particular case $\sigma/\epsilon=-0.75$,
the reflection coeficient tends to 1, reproducing NBC for $x\ll 1$.
} \label{fig1}
\end{figure}

\section{Generalization to arbitrary surfaces and worldline
approach}\label{sec:general}
In the previous sections, we have
introduced nonlocal interaction terms, which may be regarded as
`regularized' versions of the terms one should introduce to impose
Dirichlet or Neumann boundary conditions. Indeed, the singular terms
involving Dirac's $\delta$ function or its derivatives disappear, at
the price of introducing a finite length scale.

That step is not necessary for the Dirichlet case (unless one wanted
to describe a medium with an intrinsic nonlocality). On the other
hand, because of its highly singular nature, a {\em local\/} Neumann
term should be approached as a special limit of a nonlocal term.
Indeed, one should do so in order to tame the infinities that
otherwise would pop up at a very early stage, namely, when
calculating the reflection and transmission coefficients, {\em
before\/} dealing with the Casimir energy.

The explicit construction of the nonlocal terms has, however, only
been carried out for the simplest possible case, regarding the
mirror's geometry. Indeed, we have considered terms which, in the
local limit, would correspond to a plane mirror at $x_d=0$.

Let us now extend the construction to more general surfaces.  We
have in mind a situation where one really wants to deal with a
Neumann-like condition on a local (zero-width) surface and, in order
to do that, one temporarily introduces a small nonlocality along the
normal direction to the surface, on a length scale $\sim \epsilon$.
That nonlocality  is introduced to regularize the problem, and it
should disappears in the end when one takes the `zero width Neumann
limit' $\epsilon \to 0$ (and the corresponding limit for the
coupling constant). Because of this reason, we shall make use of
some simplifications that stem from the fact that, even though it is
different from zero, $\epsilon$ is very small in comparison with the
other characteristic length scales in the system.

The only assumption about the surface will be that it is piecewise
regular. Besides, it is sufficient to consider just one surface.
Indeed, for more than one surface, one just have to add more terms
to the action: one for every disconnected piece.  Also, for the sake
of simplicity, we shall first assume that the nonlocality only
exists for the normal direction to the surface.  Finally, we shall
restrict our study to the case of surfaces in three-dimensional
space.

Let us begin by writing the well-known {\em local\/} interaction
term, corresponding to Dirichlet like boundary conditions, for a
zero-width static surface $\Sigma$, and generalize it to the
nonlocal case afterwards. This surface will be assumed to be given
in parametric form:
\begin{equation}
\Sigma:\; (\sigma_1,\,\sigma_2)  \to {\mathbf Y}(\sigma) \;,
\;\;\;{\mathbf Y}(\sigma) \in {\mathbb R}^{(3)}\;\;.
\end{equation}

Then, the local Dirichlet-like interaction term $S_\Sigma(\sigma)$,
is:
\begin{equation}
S_\Sigma(\varphi) \;=\; \frac{\mu_0}{2} \,\int dx_0 d\sigma_1
d\sigma_2 \sqrt{G(\sigma)} \, \big(\varphi[x_0,{\mathbf Y}(\sigma)]
\big)^2
\end{equation}
where $G(\sigma) \equiv \det [G_{ab}(\sigma)]$, with $G_{ab}(\sigma)
\equiv {\mathbf T}_a(\sigma) \cdot {\mathbf T}_b(\sigma)$,
$a,b=1,2$, and ${\mathbf T}_a(\sigma) \equiv \partial_a {\mathbf
Y}(\sigma)$.

We then introduce a nonlocality in this term, proceeding as follows:
we first construct the finite volume region, $\Sigma_\epsilon$, that
results from `dragging' $\Sigma$ along the normal direction. The
volume $\Sigma_\epsilon$ is spanned by introduced a new parameter,
$\eta$ (to be denoted also by $\sigma_3$), in such a way that:
\begin{equation}
\Sigma_\epsilon:\; (\sigma_1,\sigma_2,\eta) \to {\mathbf
X}(\sigma,\eta) \;,
\end{equation}
such that ${\mathbf X}(\sigma,0)= {\mathbf Y}(\sigma)$, $\forall
\sigma$. $\eta$ will have infinitesimal values around zero and we
want it to introduce departures in the normal direction only. Then,
to first-order in $\eta$, a parametrization for $\Sigma_\epsilon$
can be explicitly written:
\begin{eqnarray}
 {\mathbf X}(\sigma,\eta) &=&  {\mathbf X}(\sigma,0) \,+\,
[\partial_\eta{\mathbf X}(\sigma,\eta) ]_{\eta=0} \, \eta \,+\,
{\mathcal
O}(\eta^2) \nonumber\\
 &=&  {\mathbf Y}(\sigma) \,+\, \widehat{\mathbf N}(\sigma)\, \eta \,+\, {\mathcal
O}(\eta^2) \;,
\end{eqnarray}
where we introduced the unit normal vector field:
\begin{equation}
 \widehat{\mathbf N}(\sigma)\,=\, \frac{{\mathbf T}_1(\sigma) \times
{\mathbf T}_2(\sigma)}{|{\mathbf T}_1(\sigma) \times {\mathbf
T}_2(\sigma)|} \;,
\end{equation}
at each point of the surface (which is assumed to be regular).

In these coordinates, for small $\eta$, and using the index $3$ for
$\eta$, the metric tensor $G_{ij}(\sigma,\eta)$,
($i,\,j=\,1,\,2,\,3$) on $\Sigma_\epsilon$ becomes:
$G_{ab}(\sigma,\eta) = G_{ab}(\sigma)$, $G_{aj}(\sigma,\eta) =
G_{ja}(\sigma,\eta) = 0$, and $G_{33} = 1$.

With these conventions, we may write the local term above as
follows:
\begin{eqnarray}
S_\Sigma(\varphi) &=& \frac{\mu_0}{2} \,\int dx_0 d\sigma_1
d\sigma_2 d\eta \, \sqrt{G(\sigma)} \, \delta(\eta) \,
\big(\varphi[x_0,{\mathbf X}(\sigma,\eta)] \big)^2 \nonumber\\
&=& \frac{\mu_0}{2} \,\int dx_0 d\sigma_1 d\sigma_2
\sqrt{G(\sigma)}\, \big(\varphi[x_0,{\mathbf Y}(\sigma)] \big)^2
\;,
\end{eqnarray}
as it should be.

The nonlocal term (either Dirichlet or Neumann-like) is then
constructed in a quite straightforward way:
\begin{eqnarray}
S_{\Sigma_\epsilon}(\varphi) &=& \frac{\lambda}{2} \,\int dx_0
d\sigma_1 d\sigma_2
\sqrt{G(\sigma)} \,d\eta \,d\eta' g_\epsilon(\eta) \,  g_\epsilon(\eta') \, \nonumber\\
&\times& \varphi[x_0,{\mathbf X}(\sigma,\eta)]  \;
\varphi[x_0,{\mathbf X}(\sigma,\eta')] \;,
\end{eqnarray}
where $g_\epsilon$ has the form of an approximant of the $\delta$ in
the Dirichlet case, and of its derivative in the Neumann case. Note
that, with our conventions, $\eta$ has the dimensions of a length.
Thus we may effectively assume that because of the function
$g_\epsilon$, the relevant range of $\eta$ is $\sim
[-\frac{\epsilon}{2}, \frac{\epsilon}{2}]$.

As a concrete example, we may write $S_{\Sigma_\epsilon}$ for a
sphere of radius $R$:
\begin{eqnarray}
S_{\Sigma_\epsilon}(\varphi) &=& \frac{\lambda}{2} R^2 \, \int dx_0
\, d\theta \, d\varphi \,\sin^2\theta
\,dr \,dr' g_\epsilon(r) \,  g_\epsilon(r') \, \nonumber\\
&\times& \varphi[x_0,{\mathbf X}(\theta,\varphi,r)]  \;
\varphi[x_0,{\mathbf X}(\theta,\varphi,r')] \;,
\end{eqnarray}
where we used spherical coordinates.

We conclude by presenting the implementation of this type of term
within the worldline approach to Casimir effect~\cite{Gies}, a very
useful tool for the calculation of Casimir energies.  The usual
worldline applies, by construction, to Dirichlet-like boundary
conditions, which emerge as the result of the introduction of a {\em
local\/} potential term.  On the other hand, as we have shown,
Neumann conditions require the consideration of nonlocal terms, even
when one is interested in imposing Neumann boundary conditions on a
zero-width surface.

Let us consider here the changes one has to introduce in the
worldline approach, to be able to deal with nonlocal potentials (we
have in mind just the kind of nonlocality considered in the previous
sections). For the Casimir energy of a (massive, for the sake of
completeness) field in the worldline approach, the starting point is
formally the same as in the local case; indeed, one first defines an
effective action $\Gamma[V_\epsilon]$:
\begin{equation}
\Gamma[V_\epsilon] \;=\;\frac{1}{2}\, {\rm Tr} \ln \left[
\frac{-\partial^2 + m^2 \,+\, V_\epsilon}{-\partial^2 + m^2} \right]
\end{equation}
where now $V_\epsilon$ is an operator whose matrix elements, in the
coordinate representation, are:
\begin{eqnarray}
{\mathcal V}_\epsilon (x,x') &=& \langle x | V_\epsilon |x' \rangle
\,=\, \langle x_0, {\mathbf x} | V_\epsilon |x'_0, {\mathbf x'}
\rangle
\nonumber\\
&=& \mu_0 \,\delta(x_0-x'_0) \, \int d^2\sigma \, \sqrt{G(\sigma)}
\,\int d\eta \,d\eta'
\nonumber\\
&\times& g_\epsilon(\eta) \delta^{(3)}\big({\mathbf x}-{\mathbf
X}(\sigma,\eta)\big)\, g_\epsilon(\eta') \,
\delta^{(3)}\big({\mathbf x'}-{\mathbf X}(\sigma,\eta')\big)
\end{eqnarray}
(we follow here the usual convention whereby Dirac's notation is
used for the Hilbert space of functions of $x$).

Then, using Frullani's representation \cite{frulla} for the
logarithm of a ratio, one has:
\begin{equation}
\Gamma[V_\epsilon] \;=\; -\frac{1}{2}\, \int_{0+}^\infty
\frac{dT}{T} \;\left[  K(T) - K_0(T) \right] \;,
\end{equation}
where:
\begin{equation}
K(T) \;=\; \int d^4x \, K(x,T;x,0) \;,
\end{equation}
\begin{equation}
K(x'',T;x',0) \;\equiv\; \langle x'' | e^{-T H} |x' \rangle \;,
\end{equation}
with $H \;=\; p^2 \,+ \,m^2 \,+\, V_\epsilon \;\equiv\; H_0 \,+\,
V_\epsilon$, and
\begin{equation}
K_0(x'',T;x',0) \;\equiv\; \langle x'' | e^{-T H_0} |x' \rangle \;.
\end{equation}

The only place where a departure with respect to the local case
appears is in the path integral representation for $K(T)$.
Partitioning the $T$ interval into many equal steps and evaluating
the transition amplitude for an infinitesinal evolution in each
step, one realizes that, in the limit when the number of steps tends
to infinity, the following path integral representation emerges:
\begin{equation}
K(T) \;=\; {\mathcal N} \, e^{-m^2 T} \, \int_{x(T)=x(0)} {\mathcal
D}x \, e^{- {\mathcal S}[x]}
\end{equation}
where:
\begin{equation}
{\mathcal S}[x] \,\equiv\, {\mathcal S}_0[x] \,+\, {\mathcal S}_I[x]
\end{equation}
with
\begin{equation}
{\mathcal S}_0[x] \,=\, \int_0^T d\tau \, \frac{1}{4}
\dot{x}^2(\tau) \;,
\end{equation}
\begin{equation}
{\mathcal S}_I[x] \;=\; \int_0^T d\tau \, \int_0^T d\tau' \,
{\mathcal V}_\epsilon [x(\tau),x(\tau')]\;,
\end{equation}
and ${\mathcal N}$ is a factor that comes from the path integration
over momenta, and is independent of the potential.

Finally, coming back to the effective action, and using the known
result for the free transition amplitude, one may write:
\begin{equation}
\Gamma[V_\epsilon] \;=\; -\frac{1}{2}\,
\frac{1}{(4\pi)^{\frac{d+1}{2}}} \, \int_{0+}^\infty \frac{dT}{T^{1
+ \frac{d+1}{2}}} e^{-m^2 T} \;\left[ \langle e^{- {\mathcal S}_I
[x] } \rangle   -  1 \right] \;,
\end{equation}
where
\begin{equation}
\langle e^{- {\mathcal S}_I [x] } \rangle \;=\;
\frac{\int_{x(0)=x(T)} {\mathcal D}x \;e^{-{\mathcal S}_I[x]} \,
e^{- {\mathcal S}_0[x]}}{ \int_{x(0)=x(T)} {\mathcal D}x \; e^{-
{\mathcal S}_0[x]}}\;.
\end{equation}

We do not dwell with the numerical evaluation of this kind of path
integral, which may certainly be more difficult than its local
counterpart, since the interaction term is not simply the integral
of a local function of time. However, since the scale of nonlocality
is assumed to be small, we expect the properties of the integral not
to be dramatically different to the standard ones. Finally, a
scaling to unit proper time may be of course implemented in a
similar way to the local case, as well as the extraction of a
`center of mass' for the closed paths involved in the integral over
paths.

\section{Conclusions}
We have shown that NBC can be obtained by coupling the field to an
interaction term which, if one wanted to have well-defined
transmission and reflection properties, has to include an intrinsic
regularization. We have also provided an explicit mechanism to
introduce that regularization in the calculation.

That regularization may be naturally thought of as due to two
sources: a finite width, and a finite nonlocal coupling. Neumann
conditions on a zero width surface emerge when one takes a special
limit, whereby the coupling constant is related to the width. One
may then conclude that, to reach NBC, one needs essentially just one
constant to control the UV behaviour, rather than two independent
scales. Of course, if one wanted to deal with a phenomenological
model where the nonlocal term is derived, one may obtain more that
one scale, and the resulting boundary conditions may exhibit a
richer structure.

We have presented a possible way to implement the kind of path
integral that one would require if one wanted to use a worldline
approach to nonlocal terms. We believe such an approach might shed
light on the properties of the Casimir energy with NBC for arbitrary
surfaces.
\section*{Acknowledgements}
C.D.F. thanks CONICET, ANPCyT and UNCuyo for financial support. The
work of F.D.M. and F.C.L was supported by UBA, CONICET and ANPCyT.

\end{document}